\documentstyle[12pt,emulateapj,apjfonts,psfig]{article}

\tighten
\lefthead{Paciesas et al.}
\righthead{Fourth BATSE GRB Catalog}

\begin{document}

\submitted{Accepted for publication in June 1999}

\title{The Fourth BATSE Gamma-Ray Burst Catalog (Revised)}

\author{
William S. Paciesas\altaffilmark{1,8},
Charles A. Meegan\altaffilmark{2},
Geoffrey N. Pendleton\altaffilmark{1,8},
Michael S. Briggs\altaffilmark{1},
Chryssa Kouveliotou\altaffilmark{3},
Thomas M. Koshut\altaffilmark{3},
John Patrick Lestrade\altaffilmark{4},
Michael L. McCollough\altaffilmark{3},
Jerome J. Brainerd\altaffilmark{1},
Jon Hakkila\altaffilmark{5},
William Henze\altaffilmark{6},
Robert D. Preece\altaffilmark{1},
Valerie Connaughton\altaffilmark{2,7},
R. Marc Kippen\altaffilmark{8},
Robert S. Mallozzi\altaffilmark{8},
Gerald J. Fishman\altaffilmark{9},
Georgia A. Richardson\altaffilmark{8},
Maitrayee Sahi\altaffilmark{3}
}

\altaffiltext{1}{Department of Physics, University of Alabama in
 Huntsville, Huntsville, AL 35899}
\altaffiltext{2}{NASA/Marshall Space Flight Center, ES84, Huntsville, AL
 35812}
\altaffiltext{3}{Universities Space Research Association, MSFC, ES84,
 Huntsville AL 35812}
\altaffiltext{4}{Dept. of Physics, Mississippi State University, MS
 39762}
\altaffiltext{5}{Dept. of Physics \& Astronomy, Minnesota State
 University, Mankato, MN 56002}
\altaffiltext{6}{Teledyne Brown Engineering, MSFC, ES84, Huntsville, AL
 35812}
\altaffiltext{7}{NAS$/$NRC Resident Research Associate}
\altaffiltext{8}{Center for Space Plasma, Aeronomical and Astrophysical
 Research, University of Alabama in Huntsville, Huntsville,
AL 35899}
\altaffiltext{9}{NASA$/$Marshall Space Flight Center, ES01, Huntsville, AL
35812}

\begin{center}
To appear in {\it The Astrophysical Journal Supplement Series} \\
\copyright \hspace{0.05mm} 1999 by the American Astronomical Society.
\end{center}

\begin{abstract}

The Burst and Transient Source Experiment (BATSE) on
the Compton Gamma Ray Observatory (CGRO) has triggered on
1637 cosmic gamma-ray bursts between 1991 April~19 and 1996
August~29.  These events constitute the Fourth BATSE burst
catalog. The current version (4Br) has been revised from the
version first circulated on CD-ROM in September 1997 (4B) to
include improved locations for a subset of bursts that have
been reprocessed using additional data. A significant
difference from previous BATSE catalogs is the inclusion of
bursts from periods when the trigger energy range differed
from the nominal 50--300 keV. We present tables of the burst
occurrence times, locations, peak fluxes, fluences,  and
durations.  In general, results from previous BATSE catalogs
are confirmed here with greater statistical significance.

\end{abstract}

\keywords{ gamma rays: bursts -- gamma rays:
observations -- catalogs}

\section{Introduction}

BATSE observations of gamma-ray bursts (GRBs) provided
the first clear indication of their extragalactic origin.
The angular distribution is isotropic, while the intensity
distribution shows fewer weak bursts than would be expected
from a homogeneous distribution of sources in Euclidean
space (Meegan et al.\ 1992).  No observed Galactic component
has these spatial properties. Now that some bursts have been
associated with optical counterparts that appear to be
extragalactic (van Paradijs et al.\ 1997; Metzger et al.\
1997; Sahu et al.\ 1998), it is reasonable to conclude that
all bursts come from cosmological distances.

The first catalog (1B) of BATSE bursts (Fishman et al.\
1994) consisted of 260 bursts, and covered the time interval
from 1991 April~19 until 1992 March~5. The second catalog
(2B) and third catalog (3B) extended the time interval to
1993 March~9 (585 bursts) and 1994 September~19 (1122
bursts), respectively (Meegan et al.\ 1994, 1996). We present
here the fourth catalog (4Br), which includes 1637 bursts
detected from launch through 1996 August 29. The current
version (4Br) has been revised from the version first
circulated on CD-ROM in September 1997 (4B) and described by
Meegan et al.\ (1998), to include improved locations for a
subset of bursts that have been reprocessed using additional
data. The summary tables herein include only the revised 3B
and post-3B bursts. The full catalog data are available
electronically on the World Wide Web at
http:$/$$/$www.batse.msfc.nasa.gov$/$\-batse$/$\-grb$/$\-catalog$/$\-4b$/$ or
http:$/$$/$cossc.gsfc.nasa.gov$/$\-cossc$/$\-batse$/$\-4Bcatalog$/$\-4b\_catalog
.html.

\section{Instrumentation}

BATSE consists of eight detector modules situated at
the corners of the CGRO spacecraft.  Each module contains a
50.8~cm diameter by 1.27~cm thick NaI scintillator,
sensitive to gamma-rays from $\sim$25--2000 keV.  For details of
the experiment, see Fishman et al.\ (1989).  The nominal CGRO
orbit altitude is 450~km. During the first few years of the
mission, atmospheric drag brought the altitude down to $\sim$350~km, 
and in 1994 the spacecraft was re-boosted to the nominal
altitude using an on-board propulsion system. In 1997, the
altitude was further boosted to $\sim$550~km in order to keep the
spacecraft in orbit through the next solar maximum.  A plot
of the CGRO altitude versus time is shown in Figure~1.

\noindent\parbox{8.75cm}{
\psfig{file=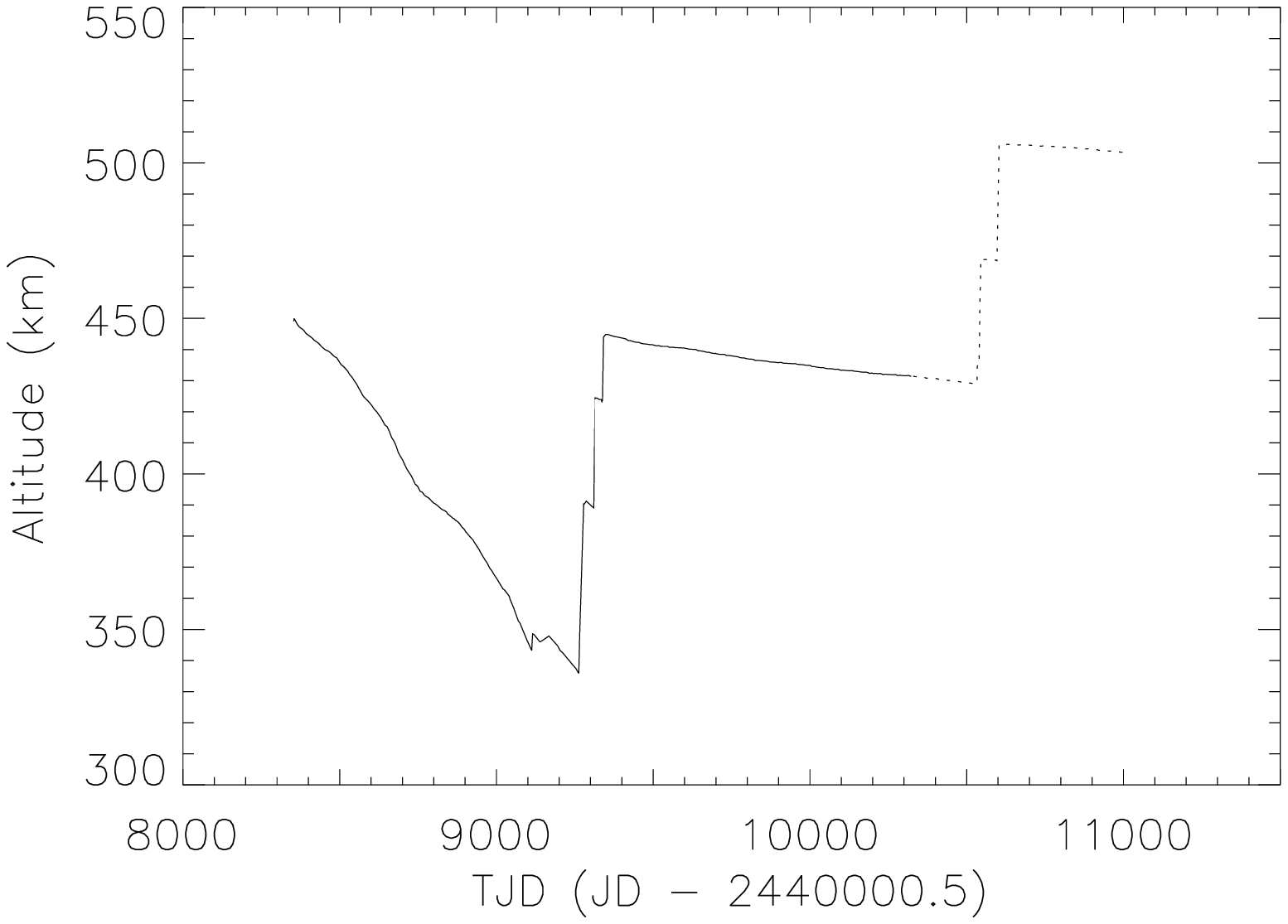,width=8.7cm,clip=}
\figcaption{CGRO altitude vs. time. The period covered by the
4Br catalog is shown by the solid curve.}
\vspace{15pt}
}

Bursts are nominally recognized on board as
simultaneous statistically significant increases, above pre-set
thresholds, in the count rates of two or more detectors
in a specified energy range.  The rates are tested at 64,
256, and 1024~ms intervals.  The background is recomputed
every 17.408~s. Prior to 1994 September~19, the trigger
energy range was set to the nominal 50--300~keV. Since then,
various scientific considerations have caused us to employ
several alternative trigger energy ranges, as summarized in
Table~1. This results in spectrum-dependent differences in
trigger sensitivity, so that the different trigger energy
settings should properly be treated as separate burst
experiments. Table~2 summarizes the total amount of time in
each trigger range and the corresponding number of bursts.
Furthermore, the levels of the three thresholds are
independently adjustable by command, specified in units of
standard deviations $\sigma$ above the background rate, and the
three trigger time scales have different sensitivities
depending on the temporal structure of the bursts, so that
each is best thought of as a separate burst experiment.  The
history of the threshold settings is also included in Table~1.

\begin{deluxetable}{rrcrrr}
\tablefontsize{\small}
\tablecolumns{6}
\tablenum{1}
\tablecaption{History of Trigger Channels \& Threshold Settings}
\tablehead{
   Dates  & TJD$/$Seconds & Channels  &
  \multicolumn{3}{c}{Thresholds ($\sigma$)}  \nl
   &  &  &    64 ms  &  256 ms & 1024 ms}
\startdata
    19 Apr 91  &               & 2+3    &   5.5   &    5.25  &     5.0 \nl
    28 Apr 91  &  8367$/$5632  &        &         &    5.5   &    5.5  \nl
    10 May 91  & 8386$/$75036  &        &         &          &   7.0   \nl
    4 Jun 91   & 8411$/$68201  &        &         &          &    5.5  \nl
    18 Aug 92  & 8852$/$56078  &        &         &          &   10.0  \nl
    24 Aug 92  & 8858$/$81762  &        &         &    8.0   &    8.0  \nl
    26 Aug 92  &  8860$/$78199 &        &         &          &    10.0 \nl
    14 Sep 92  & 8879$/$70852  &        &         &    5.5   &    5.5  \nl
    19 Sep 94  & 9614$/$57154  &   3+4  &         &          &         \nl
    31 Jan 95  & 9748$/$55085  &   1+2  &     6.0 &    6.0   &    6.0  \nl
     6 Feb 95  &  9754$/$69000 &        &    10.0 &          &         \nl
    17 Feb 95  &  9765$/$62185 &    3+4 &     4.5 &    4.5   &    4.5  \nl
    12 Apr 95  & 9819$/$56745  &   1+2  &     6.0 &    6.0   &    6.0  \nl
    10 May 95  &  9847$/$74116 &        &         &          &    10.0 \nl
    20 Jul 95  & 9918$/$73523  &        &    20.0 &   10.0   &         \nl
    21 Jul 95  &  9919$/$62439 &        &    10.0 &          &         \nl
    24 Jul 95  & 9922$/$54971  & 1+2+3+4 &   26.0 &    6.0   &    6.0  \nl
    28 Jul 95  & 9926$/$66825  &        &    10.0 &          &         \nl
   5 Sep 95   & 9965$/$60364   &        &    5.5  &   5.5    &   5.5   \nl
   2 Oct 95   & 9992$/$77028   &  1+2   &         &          &   7.0   \nl
   23 Oct 95  & 10013$/$73672  &  2+3   &         &          &   5.5   \nl
   11 Dec 95  & 10062$/$77542  &   1    &         &          &   3.5   \nl
   18 Dec 95  & 10069$/$64796  &        &         &          &   4.0   \nl
    7 Jan 96  & 10089$/$62939  & 1+2+3  &         &          &   5.5   \nl
    5 Apr 96  & 10178$/$85399  &  2+3   &         &          &         \nl
   25 Jun 96  & 10259$/$53576  &  3+4   &    4.5  &   4.5    &   4.5   \nl
   29 Aug 96  & 10324$/$77818  &  2+3   &    5.5  &   5.5    &   5.5   \nl
\enddata
\end{deluxetable}

\begin{deluxetable}{cccc}
\tablefontsize{\small}
\tablecolumns{4}
\tablenum{2}
\tablecaption{Trigger Energy Range Summary}
\tablehead{
  Channels   &   Energy (keV)   &    Days    & Number of bursts}
\startdata
      1      &      25--50      &      27     &         9    \nl
     1+2     &      25--100     &     141     &         79   \nl
    1+2+3    &      25--300     &      89     &         50   \nl
   1+2+3+4   &       $>$25    &      70     &         54   \nl
     3+4     &       $>$100   &     253     &        189   \nl
     2+3     &      50--300     &     1386    &        1256  \nl
\enddata
\end{deluxetable}

When a burst trigger occurs, BATSE enters a mode in
which high-rate data are accumulated and stored for later
transmission.  During this data accumulation interval,
further burst triggers are disabled.  The duration of the
interval, set by command, was 241.7~s from launch until 1992
July~4; then 180.2~s until 1992 July~7, then 241.7~s until
1992 Dec~17; then 573.4~s thereafter.  At the end of this
accumulation interval, a readout interval begins and the
accumulated data are transmitted.  During the readout
interval, burst triggers are enabled, but only on the 64 ms
time scale, and the trigger threshold is raised to
correspond approximately to the maximum rate of the current
burst.  A burst that triggers during this time is referred
to as an overwrite. When an overwrite occurs, the readout of
the remaining data from the overwritten event is suspended.
Consequently, some or all of the data from an overwritten
trigger may be lost.

On 1992 December~17, flight software revisions were
made to compensate for the failed CGRO tape recorders.  A
new burst data type, DISCLB, was added to enhance the
computation of burst locations even if the burst occurred
during a real-time data gap.  Also, stored commands were
initiated to suspend readout of the burst memory during
telemetry gaps,  which are predictable, and the readout
interval for weak bursts was shortened from the standard $\sim$90
minutes to $\sim$28 minutes by eliminating some high time
resolution data.  A new TDRS satellite also significantly
reduced the telemetry gaps.  In addition, the CGRO flight
software was revised on 1993 March 17 to transmit partial
BATSE data at a lower telemetry rate using the
omnidirectional antenna when TDRS coverage was not
available.  As a result of these changes, by March of 1993,
the data recovery was almost as good as before the tape
recorder failures.

\section{Instrumental Considerations}

\subsection{Trigger Criteria}

The most significant difference between the 4Br (and
4B) and the previous BATSE catalogs is that the energy range
of the burst trigger was revised several times since the end
of the 3B catalog, as summarized in Table~1. The Trigger
Criteria Table in the 4Br catalog on-line database provides
further details. Normally, all eight BATSE detectors are
enabled for burst triggering, and a trigger requires the
rates from two or more detectors to be above threshold.
However, from 1995 July~20 to 1995 July~24, the requirement
was changed to trigger if a single detector exceeded
threshold. This was an engineering test to obtain high time
resolution data on single detector phosphorescence events.
Also, from 1995 December 11--14 only 2 detectors were enabled
for burst triggering, and from 1995 December 14--18 only 4
detectors were so enabled. This was done to obtain a larger
sample of events from the bursting pulsar GRO J1744$-$28,
which was active at that time.

The variations in the trigger criteria as summarized in
Table~1 can distort some burst global properties that
involve spectral considerations, such as the burst rate,
distributions of spectral parameters, and hardness-intensity
correlations. For many studies, it may be necessary to use
bursts that have a common trigger energy range. On the other
hand, inter-comparison of samples obtained with different
trigger criteria presents new opportunities for
investigating burst properties.

\subsection{Sky Exposure}

The sky exposure is the total observing time as a
function of celestial coordinates. In computing sky
exposure, we consider Earth blockage and times that the
burst trigger is disabled. After a burst trigger, there is a
variable time interval during which the burst data are
transmitted to the ground. During this time, the trigger
thresholds are raised so that only a stronger event can
abort the current readout. Early in the mission, the burst
memory readout time interval corresponded approximately to
one satellite orbit. As described in section~2, flight
software changes were implemented subsequent to the tape
recorder failures to suspend the memory readout during
telemetry gaps and to read out only a portion of the burst
memory for weaker bursts.

In the sky exposure calculation, burst readout times
are considered dead time, so that bursts that are overwrites
should not be included in calculations that use the sky
exposure. The sky exposure is thus the total time during
which BATSE could have triggered on a burst above the
nominal threshold and is a function only of the declination
of the burst, with a dipole moment due primarily to
disabling the trigger during passages of CGRO through the
South Atlantic Anomaly, and a quadrupole moment due to Earth
blockage. Any dependence on burst right ascension averages
out over sufficiently long periods.

The algorithm used for calculating sky exposure for the
1B catalog depended on continuous data coverage and could
not be used for subsequent catalogs due to the data gaps
arising from the tape recorder failures. A new algorithm has
been developed (Hakkila et al.\ 1998a) and used to calculate
the sky exposure for each of the BATSE catalogs and subsets
of catalogs.

There are two major differences between the new
algorithm and the previous algorithm. The previous algorithm
included only time during which the trigger threshold was at
the nominal setting of 5.5$\sigma$ in each of the three trigger time
scales, while the new algorithm has no such restriction.
This change increases the 1B exposure by $\sim$14\%. The new
algorithm handles the SAA passages more accurately than the
1B calculation, which overestimated the time spent in the
SAA and therefore overestimated the dipole moment of the
exposure. Correcting for sky exposure, the full sky burst
rate above the BATSE threshold, triggering on 50--300 keV, is
$\sim$666 bursts$/$year. Figure~2 shows the exposure vs. declination
for the full 4Br catalog as well as for two important data
subsets: the times when the trigger energy channels were
channels 1+2 (25--100 keV) and 3+4 ($>$100 keV).

Table~3 indicates the average exposures, dipole
moments, and pertinent quadrupole moments of the sky
exposures shown in Figure~2. The exposures have produced
similar anisotropies (due primarily to Earth blockage) in
all published BATSE burst catalogs.

Table~4 lists the rates of bursts detected above
BATSE's minimum detection threshold for the three most
commonly-used trigger criteria, corrected for sky exposure.
The burst rates are significantly higher in trigger channels
2+3 than they are in channels 1+2 or 3+4. The high signal-to-noise
ratios in channels 1 and 4 suggest that enhanced
channel 2+3 rates are not entirely due to instrumental
effects; it appears that more observable bursts exist in the
50 to 300 keV energy range (channels 2+3) than at other
energies. This is consistent with the finding by Harris \&
Share (1998) that few extremely hard bursts exist that could
not be detected by BATSE. The spectral response of the BATSE
detectors will be discussed in more detail in section 3.3.

The exposure-corrected 2+3 burst rate can be compared
to the exposure-corrected burst rate obtained by the
untriggered burst search (J. Kommers, private
communication). These rates indicate a burst distribution
that continues to decline in number below the BATSE trigger
threshold. There is no time-dependence to the BATSE exposure-corrected
gamma-ray burst rate; this has remained roughly
constant throughout the CGRO mission.

\noindent\parbox{8.75cm}{
\psfig{file=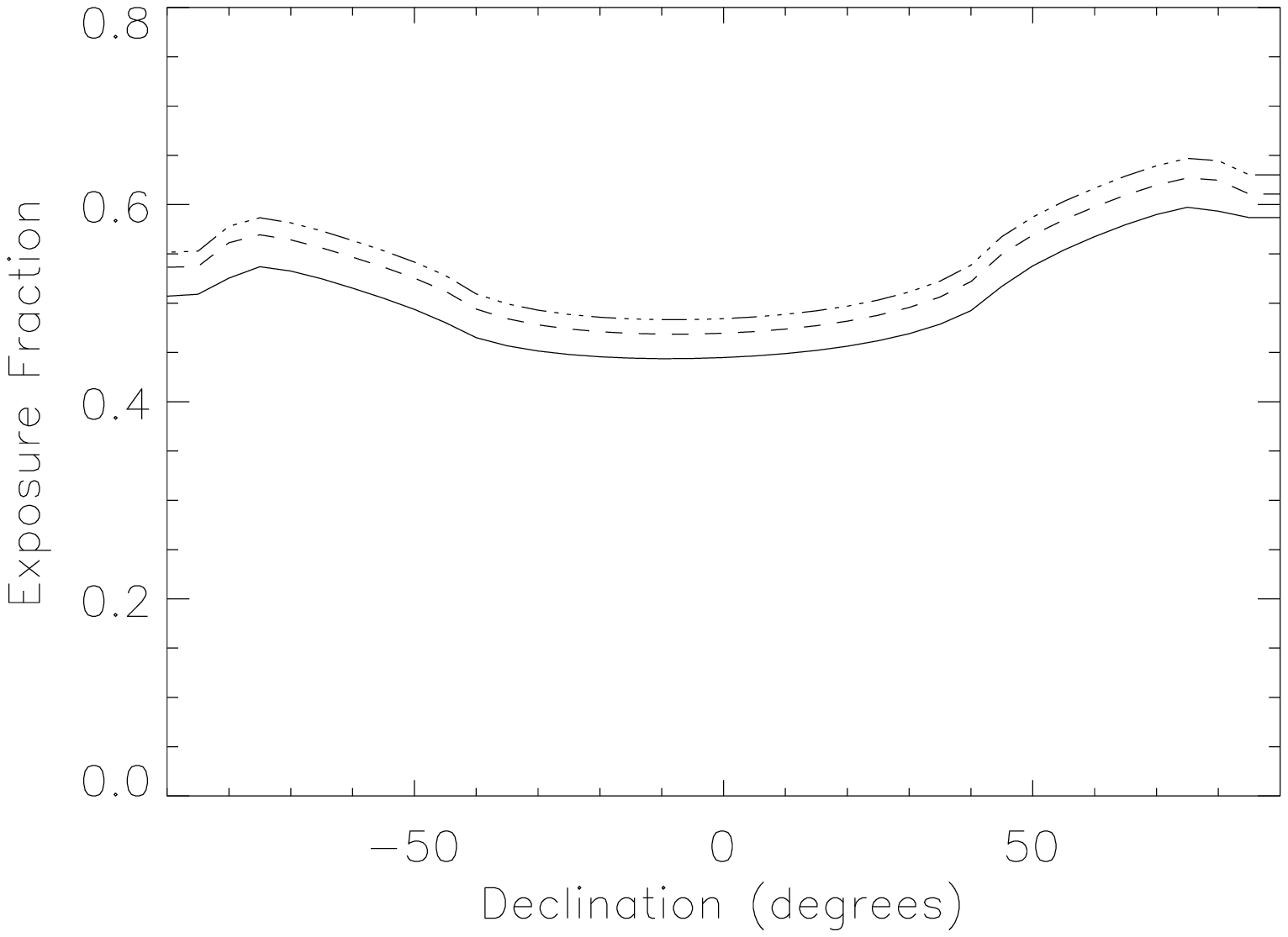,width=8.75cm,clip=}
\figcaption{Sky exposure as a function of declination for the
4Br catalog (solid curve). Additional curves show the
exposure for subsets divided according to trigger energy
range: 25--100 keV (dashes) and $>$100 keV (dot-dashes). The
exposure for the subset of 50--300 keV triggers is
indistinguishable from that of the full catalog.}
\vspace{15pt}
}

\subsection{Trigger Efficiency}

The trigger efficiency is the probability that a burst
of a given peak flux will exceed the BATSE trigger
threshold. Again, the algorithm used for the previous
catalogs cannot be used if there are data gaps. Also, the
older algorithm did not consider the increase in efficiency
due to atmospheric scattering of photons into the detectors,
the effect of the range of burst spectral properties, or the
effect of non-nominal thresholds. An improved algorithm that
overcomes these limitations is currently under development
(Pendleton, Hakkila \& Meegan 1998). Figure~3a compares
results obtained with the old and new algorithms for the
probability of exceeding the 1024~ms threshold, in the
nominal 50--300 keV trigger energy range and at the nominal
level of 5.5$\sigma$, as a function of peak flux. For this
calculation, all bursts were assumed to have the same
spectrum, viz., a typical Band function (Band et al.\ 1993).
The greater efficiency near threshold is due to inclusion of
atmospheric scattering in the new calculation. Figure~3b
compares the triggering efficiency for three different
trigger energy ranges (20--100 keV, 50--300 keV, and $>$100
keV), using the same assumed burst spectrum in all cases.

The spectral dependencies (summarized by the
$\alpha$, $\beta$, and $E_{\rm p}$
parameters of the Band function) appear to strongly
influence the trigger efficiency. The $E_{\rm p}$ parameter is
relatively important, as is the low-energy spectral index $\alpha$.
The efficiency is less sensitive to the high-energy spectral
index $\beta$. The effects of the dependence on $E_{\rm p}$ will be
demonstrated in section 4.2. As a result of these spectral
dependencies, calculation of the trigger efficiency depends
on an assumed distribution of spectral shapes, and is
therefore model-dependent.

\section{The Catalog}

The 4Br catalog includes the time period of the 3B
catalog plus an additional 515 bursts between 1994 September~20 
and 1996 August~29. The 4B catalog was initially released
in September 1997 on the World Wide Web and on CD-ROM
(Meegan et al.\ 1998). A few revised entries for 3B bursts
were included in the initial release: revised locations for
triggers 741, 2311, and 3155; a corrected date for trigger
1694; revised CMAXMIN entries for triggers 111, 3118, and
3137; and a revised duration for trigger 148. Burst
locations were computed using the same algorithm as was used
for the 3B catalog. As will be described in section 4.1, the
accuracy of the algorithm is improved substantially for
certain bursts by fitting to data from 6 detectors rather
than 4 detectors. The 4Br catalog is expanded from the
September 1997 release by including revised locations for
208 bursts (199 of these were recomputed using the same data
type, but fitting to 6 detectors, and the remaining 9 were
recomputed using more appropriate choices of data type
and$/$or time interval).

\begin{figure*}
\hfill\mbox{
\psfig{file=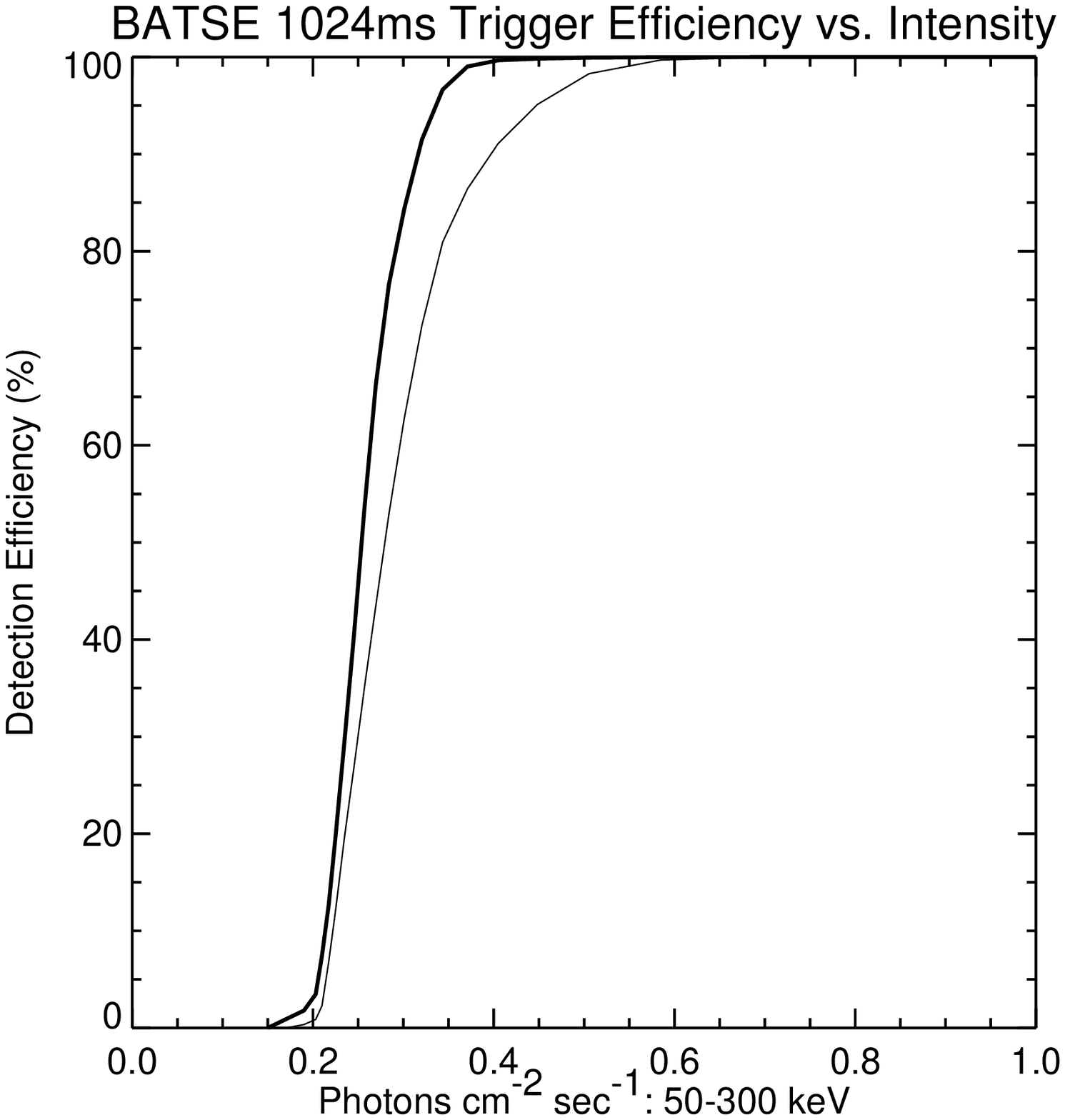,height=7.5cm,bbllx=56pt,bblly=128pt,bburx=525pt,%
bbury=600pt,clip=}
\hspace{1cm}
\psfig{file=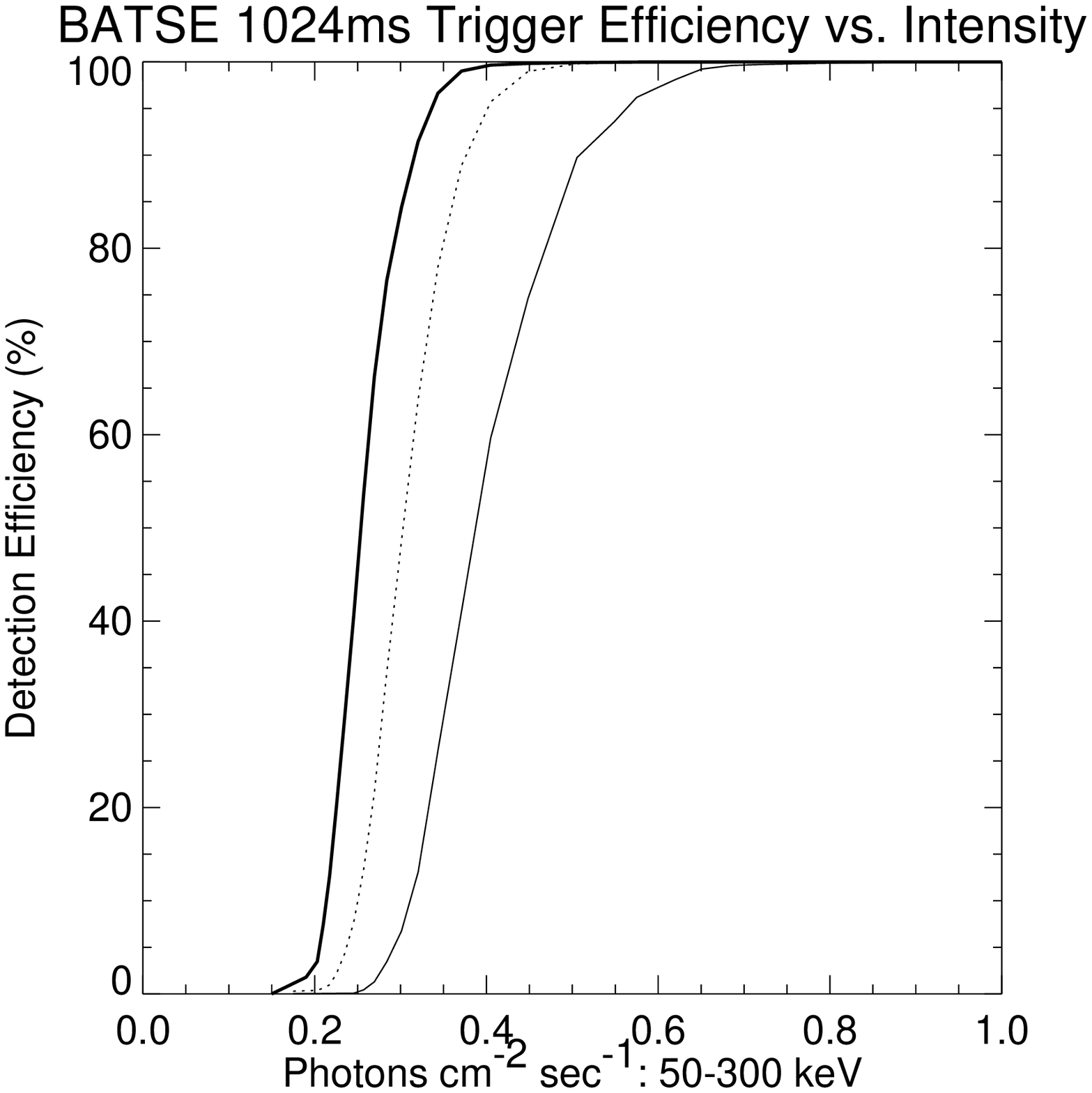,height=7.5cm,bbllx=60pt,bblly=130pt,bburx=550pt,%
bbury=597pt,clip=}}\hfill
\vspace{15pt}
\caption{\small Trigger efficiency as a function of peak flux (in
the 50--300 keV range). a) Comparison of the results using
the current (thick curve) and previous (3B catalog; thin
curve) algorithms for the nominal 50--300 keV trigger energy
range. b) Comparison of the results using the current
algorithm for different trigger energy ranges: 25--100 keV
(dotted curve), 50--300 keV (thick solid curve, the same as
in Figure 3a), and $>$100 keV (thin solid curve).}
\end{figure*}

\begin{deluxetable}{cccc}
\tablefontsize{\small}
\tablecolumns{4}
\tablenum{2}
\tablecaption{Trigger Energy Range Summary}
\tablehead{
  Channels   &   Energy (keV)   &    Days    & Number of bursts}
\startdata
      1      &      25--50      &      27     &         9    \nl
     1+2     &      25--100     &     141     &         79   \nl
    1+2+3    &      25--300     &      89     &         50   \nl
   1+2+3+4   &       $>$25    &      70     &         54   \nl
     3+4     &       $>$100   &     253     &        189   \nl
     2+3     &      50--300     &     1386    &        1256  \nl
\enddata
\end{deluxetable}

\begin{deluxetable}{cccc}
\tablefontsize{\small}
\tablecolumns{4}
\tablenum{3}
\tablecaption{Sky Exposure Summary}
\tablehead{     &  Channels 1+2 & Channels 2+3 & Channels 3+4}
\startdata
 Average relative exposure  &
                       0.508    &     0.480    &    0.524  \nl
$\langle \sin \delta \rangle$
                 &     0.017    &     0.018    &    0.017  \nl
$\langle \cos \theta \rangle$
                 &    $-$0.008    &    $-$0.009    &   $-$0.008  \nl
$\langle \sin^2 \delta - 1/3 \rangle$
                 &     0.024    &     0.024    &    0.024  \nl
$\langle \sin^2 b - 1/3 \rangle$
                 &    $-$0.004    &    $-$0.004    &   $-$0.004  \nl
\enddata
\end{deluxetable}

The 162 revised 3B bursts are listed in Table~5 and the
515 post-3B bursts are listed in Table~6. The columns are
the same as were used in the previous catalogs.  The first
column is the BATSE trigger number.  The next column
specifies the trigger name in the format 4B {\it yymmdd}, where {\it yy}
is year, {\it mm} is month, and {\it dd} is day of month.  The trigger
name may have a letter appended if there is more than one
trigger in a day.  The next column specifies the time of the
trigger expressed as the Truncated Julian Day (TJD) and the
seconds of day (s).  The next column specifies the trigger
time in the format day of year and UT time.  The next five
columns give the computed locations in equatorial (epoch
2000) and Galactic coordinates and the error in the
location. The tabulated error radius represents the radius
of a circle that has the same area as the 68\%  statistical
confidence region.  The actual 1$\sigma$ contours are not
necessarily circular.  These radii represent errors due to
photon counting statistics only; there is also a systematic
error (see section 4.1).  The next column specifies the
largest of the three values of $C_{\rm max}/C_{\rm min}$,
the maximum count rate
divided by the threshold count rate.  The next two columns
specify the threshold number of counts $C_{\rm min}$
and the relevant
trigger time scale.  Note that the latter is not necessarily
the time scale for the trigger, but the time scale for
which $C_{\rm max}/C_{\rm min}$
is the largest. The column labeled $T_{90}$ is a measure of the
burst duration.  It is the time during which  the burst
integrated counts increases from 5\% to 95\% of the total
counts.  The next column specifies the peak flux and error.
The energy interval for the peak flux is 50--300 keV (the
burst trigger range), and the integration time is 256~ms.
The next column specifies the fluence and error in the
50--300 keV energy range.  The next column presents the
hardness ratio, defined as the ratio of fluence in the
100--300 keV range to the fluence in the 50--100 keV range.
The  next column specifies the total fluence (above $\sim$20~keV)
and error over the duration of the burst.  The last column
contains codes for specific comments listed at the end of
the table.

The  $V/V_{\rm max}$ statistic (Schmidt , Higdon \& Hueter 1988) is
simply $(C_{\rm max}/C_{\rm min})^{-3/2}$.
Table~7 summarizes the average value of $V/V_{\rm max}$ for the
4Br catalog and various subsets. It is clear from the table
that none of the subsets are consistent with the value 0.5
that a homogeneous sample would produce.

The numerous missing entries are due to data gaps in
one or more of the various data types that BATSE transmits.
The various parameters have different requirements on data
completeness.  $C_{\rm max}/C_{\rm min}$
is most sensitive to missing data; durations,
fluxes, and
fluences less so.  Since locations can be computed using
several different data types and
time intervals, they are available for all bursts.

\subsection{Locations}

The 4Br catalog incorporates 211 location changes from
the 3B catalog (Meegan et al.\ 1996), of which 3 had already
been made in the 4B catalog. While twelve of the changes
were made for miscellaneous reasons, 199 of the changes are
due to relocating using the data from six instead of four
detectors. When a location obtained using the data of four
detectors has a detector with an angle larger than 90$^\circ$, this
is an indication that the choice of four detectors is poor,
since the geometry ensures that there will be always be four
detectors with source angles no greater than 90$^\circ$. We
postulated that such cases are due primarily to systematic
errors in detector response and atmospheric scattering
corrections, so that a better location would be obtained by
fitting the data from six detectors. The idea was tested by
comparing locations for 39 bursts from the 4Br catalog in
this class for which interplanetary network (IPN) annuli are
available (Hurley et al.\ 1998a,b). Two intersecting annuli
have been derived for trigger 1121, so that its true
location is precisely known. In this case, the six-detector
BATSE location is better than the four-detector case by
0.4$^\circ$, versus a statistical error of 0.3$^\circ$. For the remaining
38 events, only single IPN annuli are available, so we can
only determine the closest approach angle $\rho$ of the annuli to
the BATSE locations. Figure~4 summarizes the differences $\Delta \rho$
between the four- and six-detector locations. While not
every case shows an improvement, the average change is to a
significantly better location.

\noindent\parbox{8.75cm}{
\vspace{15pt}
\hfil\psfig{file=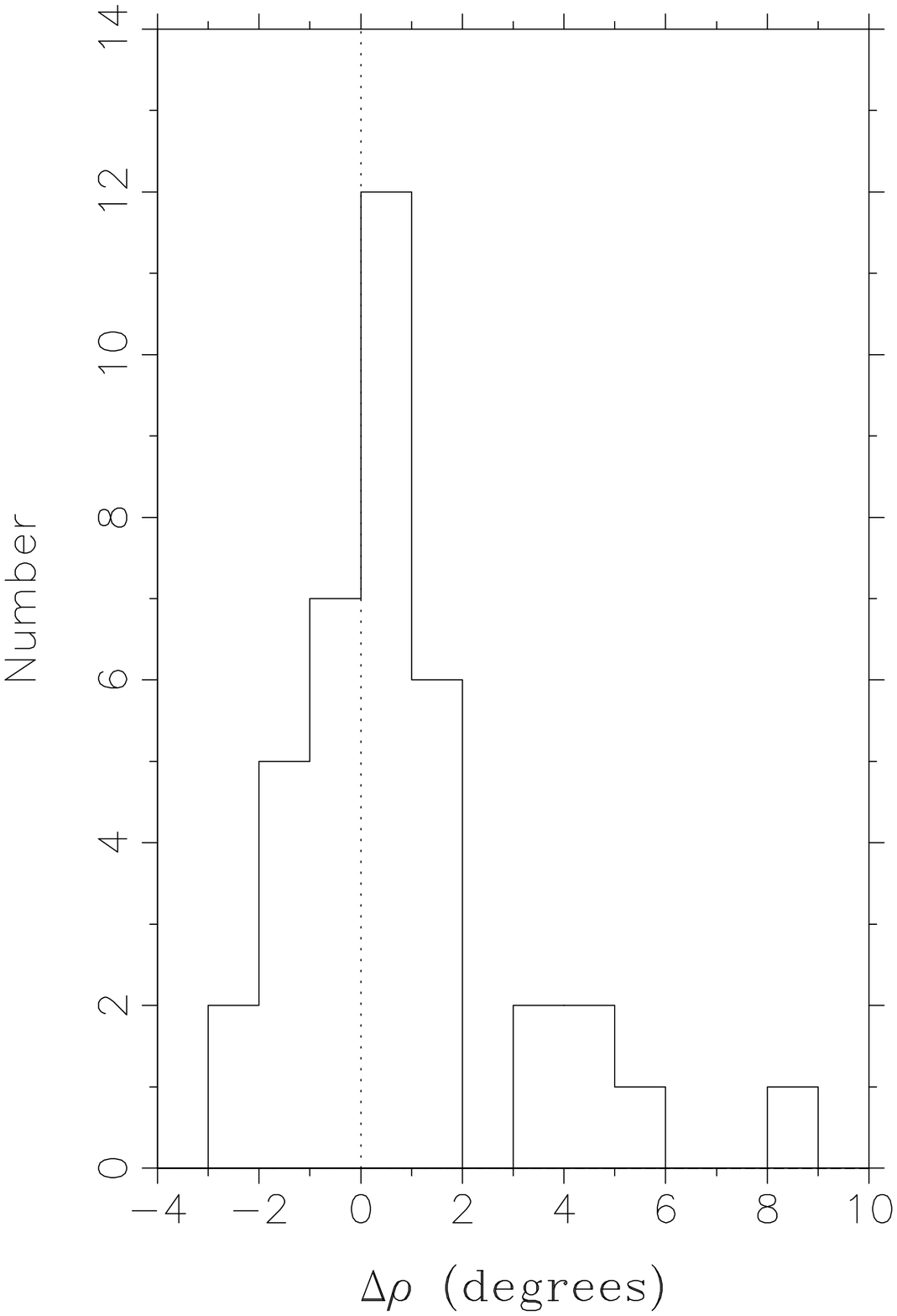,height=10cm,clip=}\hfill
\figcaption{Changes  in the distances $\Delta \rho$ of 38 BATSE locations
from single IPN annuli. A positive value of $\Delta \rho$ indicates a
location improvement. The mean improvement is 0.74$^\circ$.}
}
{
\setlength{\tabcolsep}{.2pc}
\newcommand{\nd}{\nodata}

}
\clearpage

\begin{deluxetable}{cccc}
\tablefontsize{\small}
\tablecolumns{4}
\tablenum{7}
\tablecaption{$V/V_{\rm max}$ for Various Subsets}
\tablehead{
   Subset   &   Trigger Energy  &      Number of   &  $V/V_{\rm max}$ \nl
            &       (keV)       &        bursts    &     }
\startdata
 4Br, channels 1+2      &   25--100     &     43     &     0.357$\pm$0.045   \nl
 4Br, channels 2+3      &   50--300     &     740    &     0.324$\pm$0.011   \nl
 4Br, channels 3+4      &  $>$ 100    &      103   &      0.345$\pm$0.029  \nl
 4Br, channels 1+2+3+4  &  $>$ 25     &     25     &     0.398$\pm$0.061   \nl
    Full 3B             &   50--300     &     658    &     0.329$\pm$0.011   \nl
    Full 4Br            &   various   &     911    &      0.330$\pm$0.010  \nl
\enddata
\end{deluxetable}

Tables 5 \& 6 list only the statistical location errors $\sigma_{\rm stat}$
as determined by LOCBURST (the BATSE software for
determining burst locations) from counting statistics. In
addition to $\sigma_{\rm stat}$, the location errors also include a systematic
component  $\sigma_{\rm sys}$.
In the 1B catalog, comparison of the LOCBURST
locations with more precise, independently determined,
locations for a small number of events indicated that
$\sigma_{\rm sys} = 4^\circ$
(Meegan et al.\ 1994). After improvements to the LOCBURST
algorithm, this was reduced in the 3B catalog to $\sigma_{\rm sys} = 1.6^\circ$
(Meegan et al.\ 1996). In the 3B catalog paper, it was noted that the
error distribution might not be Gaussian and that some
bursts might fall into an extended non-Gaussian tail with
higher values of $\sigma_{\rm sys}$. Recently, Briggs et al.\ (1998) used
IPN data for 411 GRBs (Hurley et al.\ 1998b) of the 4Br
catalog to test various models for the BATSE location error
distribution and to determine the optimum parameter values
of these models. An excellent fit to the data is provided by
a model in which the error distribution is a modified
Gaussian with an extended tail: 78\% of the time the
systematic error belongs to the core with a value of 1.85$^\circ$,
while the remainder of the probability corresponds to a tail
with systematic error of 5.1$^\circ$. A more complex model, in which
the systematic error depends on the data type used in
LOCBURST to obtain the location, is modestly favored by the
data. Further details and instructions for implementing the
improved error models are presented by Briggs et al.\ (1998).

Figure~5 shows the distribution in Galactic coordinates
of the locations of the 1637 bursts in the 4Br catalog. The
dipole and quadrupole moments of the observed distribution
of bursts are listed in Table~8, together with the values
expected for an isotropic distribution after correction for
sky exposure (Section 3.2). The error bars on the observed
burst moments are from the sample statistics; the location
errors make a negligible contribution. The error bars on the
sky exposure are very rough estimates. After correction for
the anisotropic sky exposure, the galactic moments and the
coordinate system independent tests (Briggs 1993) are
consistent with isotropy. Without correction, the equatorial
quadrupole moment differs from zero by more than 3$\sigma$; after
correction, both equatorial moments are consistent with
isotropy.

\begin{figure*}
\hfil\psfig{file=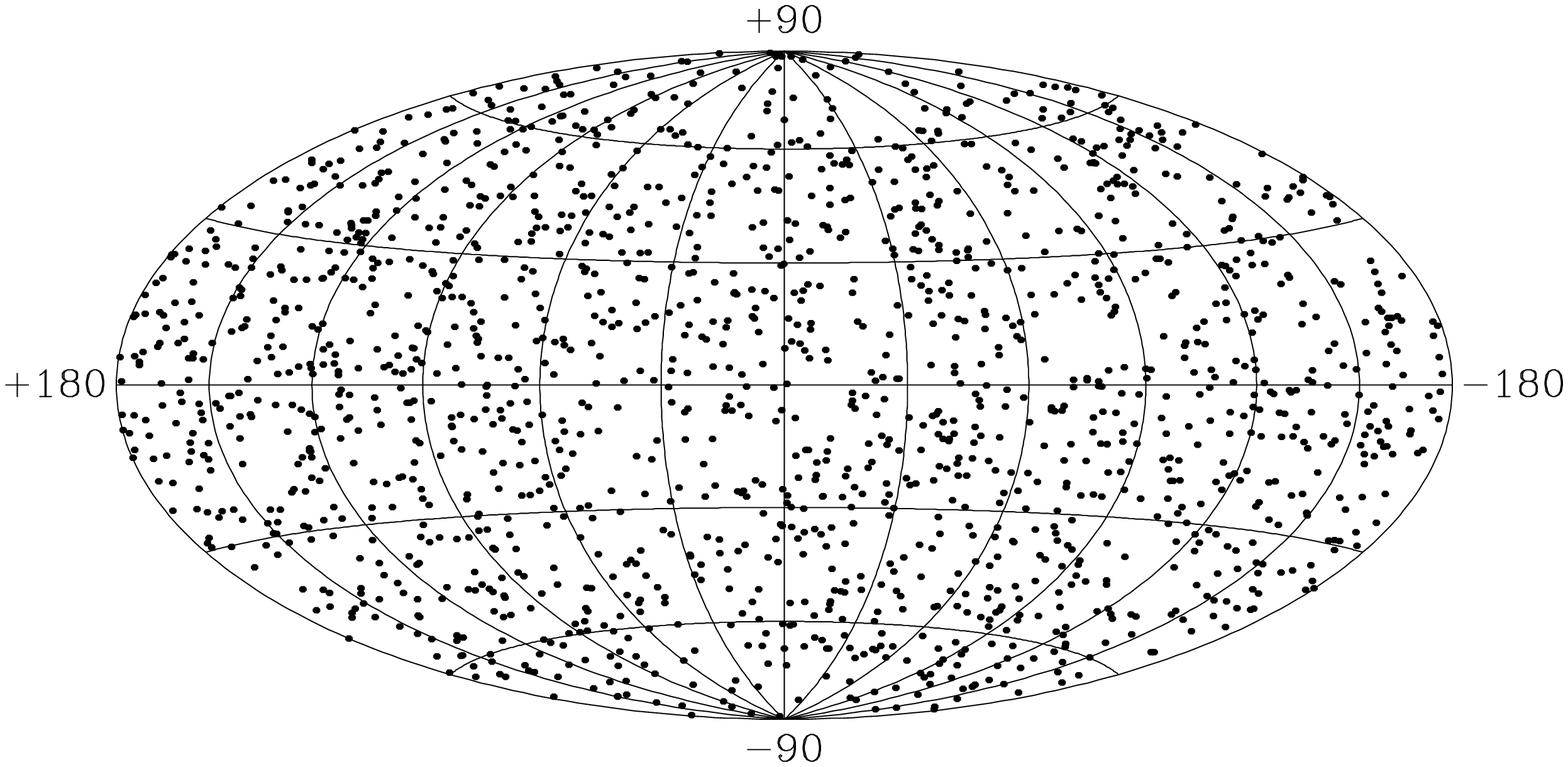,height=7.5cm,clip=}\hfill
\caption{\small Sky distribution of the 1637 bursts in the 4Br
catalog on an Aitoff-Hammer projection in Galactic
coordinates.  There is no correction for non-uniform sky
coverage. Note that the full set of 4Br locations has been
used, regardless of trigger energy range.}
\end{figure*}

We place limits on burst repetition by comparing the $R_*$
and $\langle w \rangle$ clustering statistics to Monte Carlo
simulations of the
gamma-ray burst sky distribution.  The $R_*$ statistic (Tegmark
et al.\ 1996) is a measure of burst separation weighted by
the stated burst error and a 1.6$^\circ$ systematic error,
while $\langle w \rangle$
is the two point correlation function averaged over burst
separation values of 0$^\circ$ to 5.73$^\circ$. Of these two statistics,
$R_*$ produces somewhat stronger limits on burst repetition
(Brainerd \& Kippen 1998). In the simulations, we generated
10$^4$ gamma-ray burst sky distributions with the observed
distribution of location errors for each model of a specific
number of observed repetition sources.  The simulations use
location error model 2 of Briggs et al.\ (1998), and they
take into account the sky exposure. From the simulations,
the expectation values for no repeaters are $R_* = 0.428 \pm 0.007$
and $\langle w \rangle = 0.007 \pm 0.018$,
consistent with the 4Br catalog values $R_* = 0.422$ and
$\langle w \rangle = -0.007$.  Upper limits
to the repeater fraction depend on the assumed number of
repetitions per source (Hakkila et al.\ 1998b). For models in
which repeating sources contribute 2 gamma-ray bursts to the
catalog, the repeater fraction limits at the 5\% significance
level are 8.4\% using $R_*$ and 12.9\% using $\langle w \rangle$. At the 1\%
significance level, the corresponding limits are 17.1\% using $R_*$
and 24.2\% using $\langle w \rangle$.

\subsection{Fluxes and Fluences}

Peak fluxes for each of the three trigger time scales
are determined as in the previous BATSE catalogs.  Peak
count rates are converted to units of photons cm$^{-2}$ s$^{-1}$ using
detector response matrices that include the effects of
varying angles to bursts, detector efficiency, atmospheric
scattering, and spectral response. Further details of the
method and results are presented by Pendleton et al.\ (1996).

Peak flux is here defined as the maximum flux in
photons cm$^{-2}$ s$^{-1}$, integrated over 50--300 keV in energy, and
integrated over 64~ms, 256~ms, or 1024~ms. Specifying the
peak flux in the trigger energy range provides an intensity
measurement for which instrument sensitivity can be more
directly calculated.

\begin{deluxetable}{cccccc}
\tablefontsize{\small}
\tablecolumns{6}
\tablenum{8}
\tablecaption{Dipole and Quadrupole Moments}
\tablehead{
Moment  &  Type  &   Coordinates  &   Observed &
Expected\tablenotemark{\dagger} & Deviation($\sigma$) }
\startdata
$\langle \cos \theta \rangle$
        &  dipole     & galactic    & $-$0.025$\pm$0.014  & $-$0.009  &  $-$1.1 
\nl
$\langle \sin^2 b - 1/3 \rangle$
        &  quadrupole & galactic    &  $-$0.001$\pm$0.007 &  $-$0.004 & +0.4 \nl
Watson  &  dipole     & independent &      4.0$\pm$3.6  &   4.5   &  $-$0.1 \nl
Bingham &  quadrupole & independent &     12.4$\pm$7.4  &   15.8  &  $-$0.5 \nl
$\langle \sin \delta \rangle$
        &  dipole     & equatorial  &   0.024$\pm$0.014 &  0.018  &  +0.4 \nl
$\langle \sin^2 \delta - 1/3 \rangle$
        &  quadrupole & equatorial  &   0.025$\pm$0.007 &  0.024  &  +0.1 \nl
\enddata
\tablenotetext{\dagger}{Isotropic distribution, corrected for non-uniform sky
coverage}
\end{deluxetable}

The anisotropic response of the detectors implies that
location inaccuracies produce systematic errors in
determining peak flux and fluence. However, the effect is
negligible compared to other systematic errors because the
energy dependence of the detector response changes rather
slowly with the angle of incidence, and because the use of
multiple detectors tends to average out the angle
dependence. The peak flux and fluence measurements presented
here were derived using the original 4B locations. We
verified that use of the 4Br locations would have a
negligible effect by re-computing flux$/$fluence values for 10
bursts with the largest differences between 4Br and 4B
locations. We find that the flux and fluence values are
changed by less than 10\% as a result of the location
differences. This is substantially smaller than systematic
errors due to energy calibration errors, assumed spectral
forms, and detector response uncertainties, which are
estimated to be $\sim$20\%. Figures 6a-6c show the cumulative
distributions of peak flux calculated on the 64~ms, 256~ms,
and 1024~ms trigger time scales, respectively, for bursts
observed when the trigger energy range was set to the
nominal 50--300 keV range. Near trigger threshold each of
these distributions diverges into three distinct branches,
illustrating how the instrument trigger threshold influences
the measurement of the GRB population intensity distribution
near threshold. Our continuing studies of BATSE's burst
sensitivity using the enhanced sky map algorithm (Hakkila et
al.\ 1998a; Pendleton, Hakkila \& Meegan 1998) combined with
fits of Band's spectral function (Band et al.\ 1993) to a
large population of bursts (Mallozzi et al.\ 1995) have shown
that BATSE's sensitivity to bursts is strongly dependent on
the burst
spectrum\footnote[1]{Specifically on
$E_{\rm p}$, the energy for which
$\nu F_{\nu}$ is a maximum.},
and that no single spectrum produces a
well-behaved instrument sensitivity correction near
threshold. To illustrate this, we calculated the BATSE
sensitivity for several representative spectra using
spectral parameters derived from fits of the Band function
to actual burst data. These are indicative of the range of
sensitivity corrections, but are not meant to be definitive,
as a full exploration of spectral parameter space is beyond
the scope of the present paper.

\begin{figure*}
\mbox{
\psfig{file=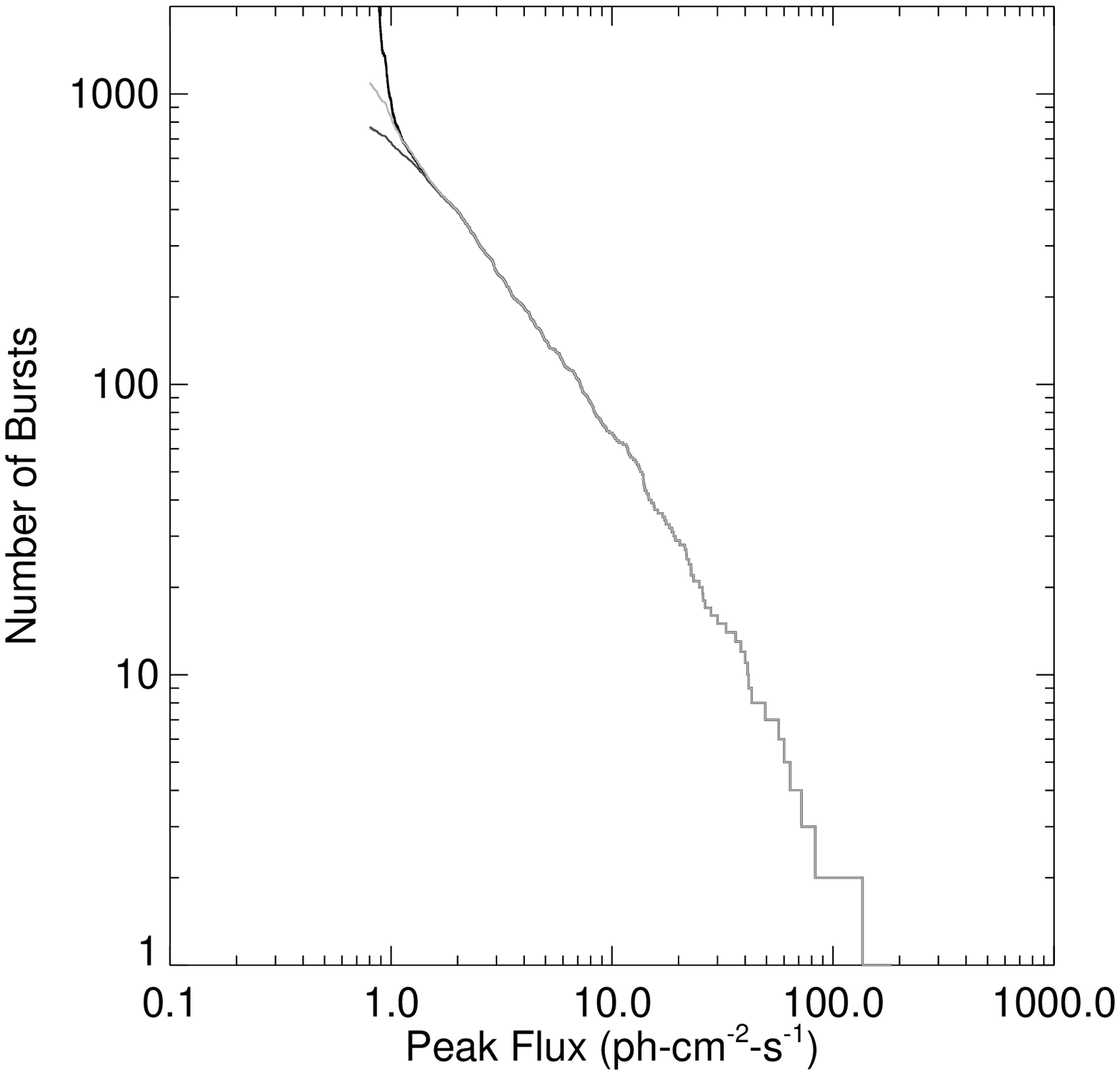,width=6cm,
clip=}
\hfill
\psfig{file=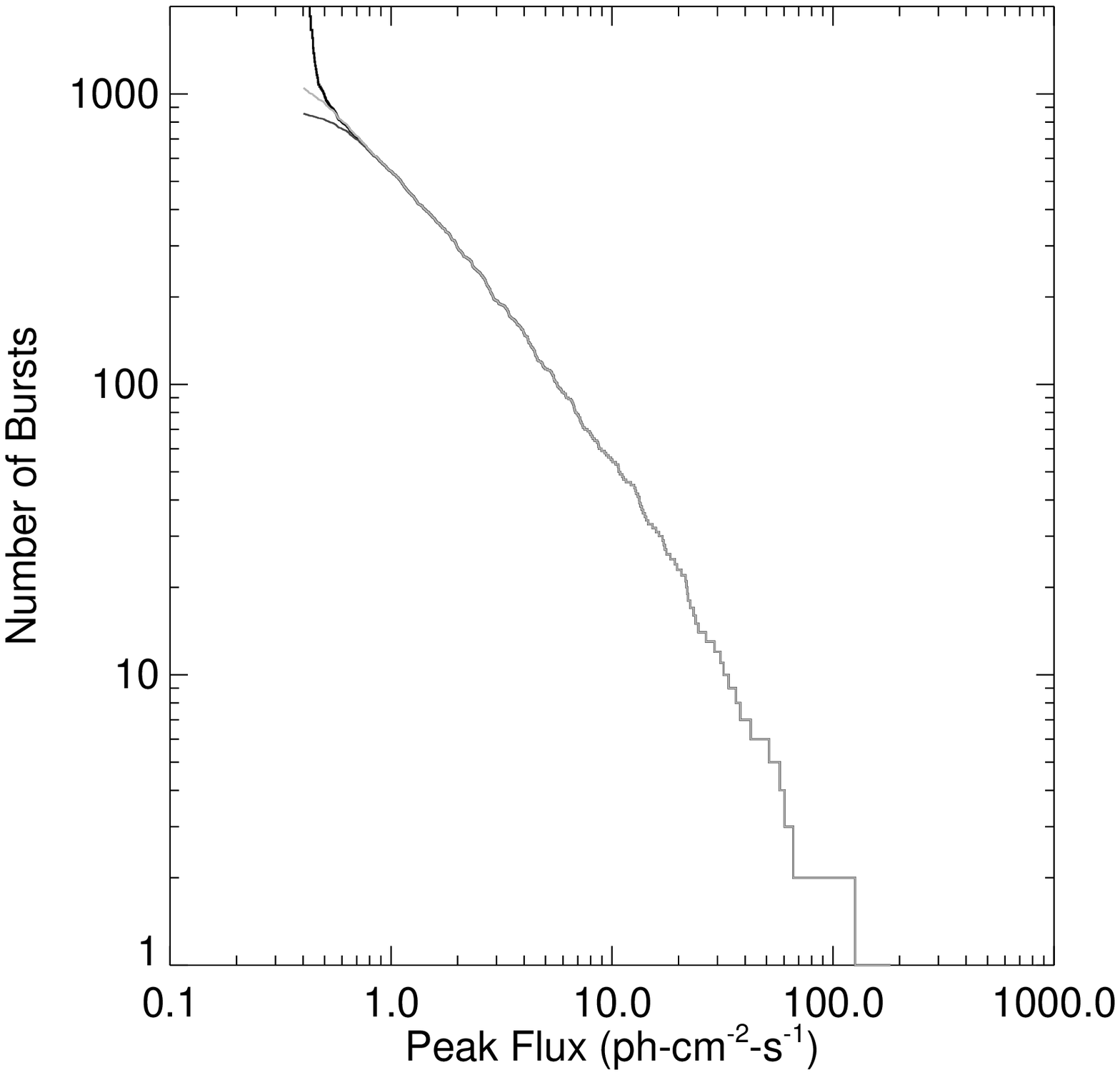,width=6cm,
clip=}
\hfill
\psfig{file=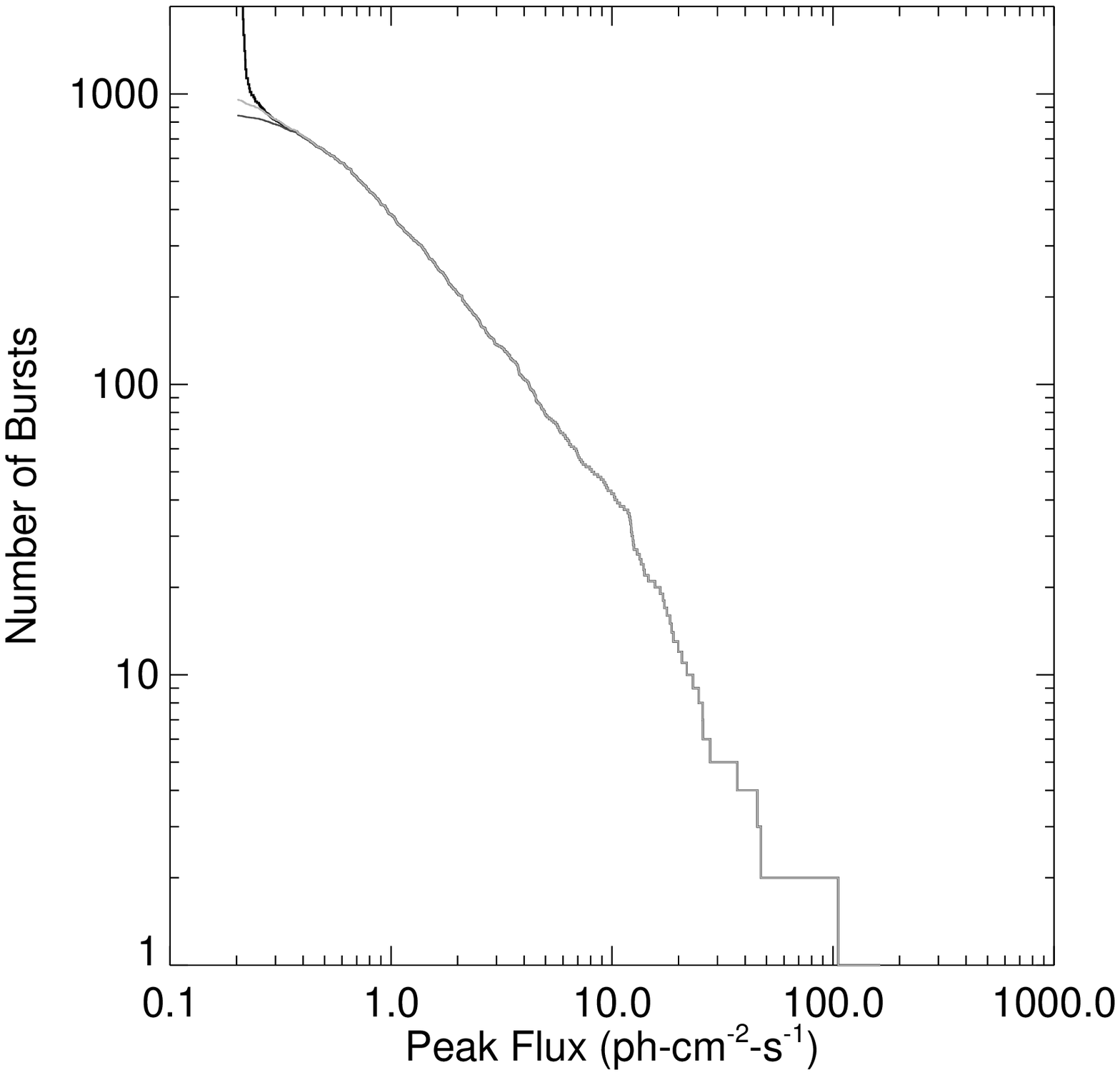,width=6cm,clip=}}\par
\caption{\small Integral $\log N - \log P$ distributions for the standard 50--300
keV trigger energy range. Each figure shows only events
above threshold on the designated time scale: a) 64 ms, b)
256 ms, c) 1024 ms. Near threshold three histograms are
shown that represent the corrections for instrumental
trigger efficiency for a range of incident spectra (see text
for details).}
\end{figure*}

In Figure~6a, the highest curve shows the peak flux
distribution using a burst sensitivity correction calculated
for a single Band function fit to a burst with $E_{\rm p}=267$ keV. For
this spectral shape, BATSE's sensitivity is relatively poor
for bursts with peak flux just above threshold. Conversely,
the lowest curve in Figure~6a shows the effect of the
sensitivity calculated using a spectrum with $E_{\rm p}=1391$ keV, for which
essentially no correction to the observed data is required.

All the sensitivity corrections calculated with
individual spectra produced corrections to the burst
intensity distribution that are relatively smooth and span a
small interval in intensity. If we define a peak flux $P_0$ above
which the sensitivity correction is negligible, then $P_0$ and
$E_{\rm p}$
are anti-correlated. Since we know that bursts exhibit a
fairly broad range of $E_{\rm p}$ for all intensities studied to date
(Mallozzi et al.\ 1995), it is reasonable to assume that a
more accurate corrected burst intensity distribution will be
obtained when a distribution of burst spectra are used in
the sensitivity calculation. The thin line in Figure~6a
shows the distribution corrected using five spectra in the
sensitivity calculation with different $E_{\rm p}$ values distributed
over the range observed by BATSE in the entire burst
population. This correction more accurately characterizes
the threshold effects than a correction obtained using a
single spectrum. It is important to emphasize, however, that
inferences concerning the ``true'' burst population statistics
for weaker bursts are quite dependent on the assumed
spectral parameters. More detailed studies are required to
quantify these effects more accurately.

\begin{figure*}
\hfil\mbox{
\psfig{file=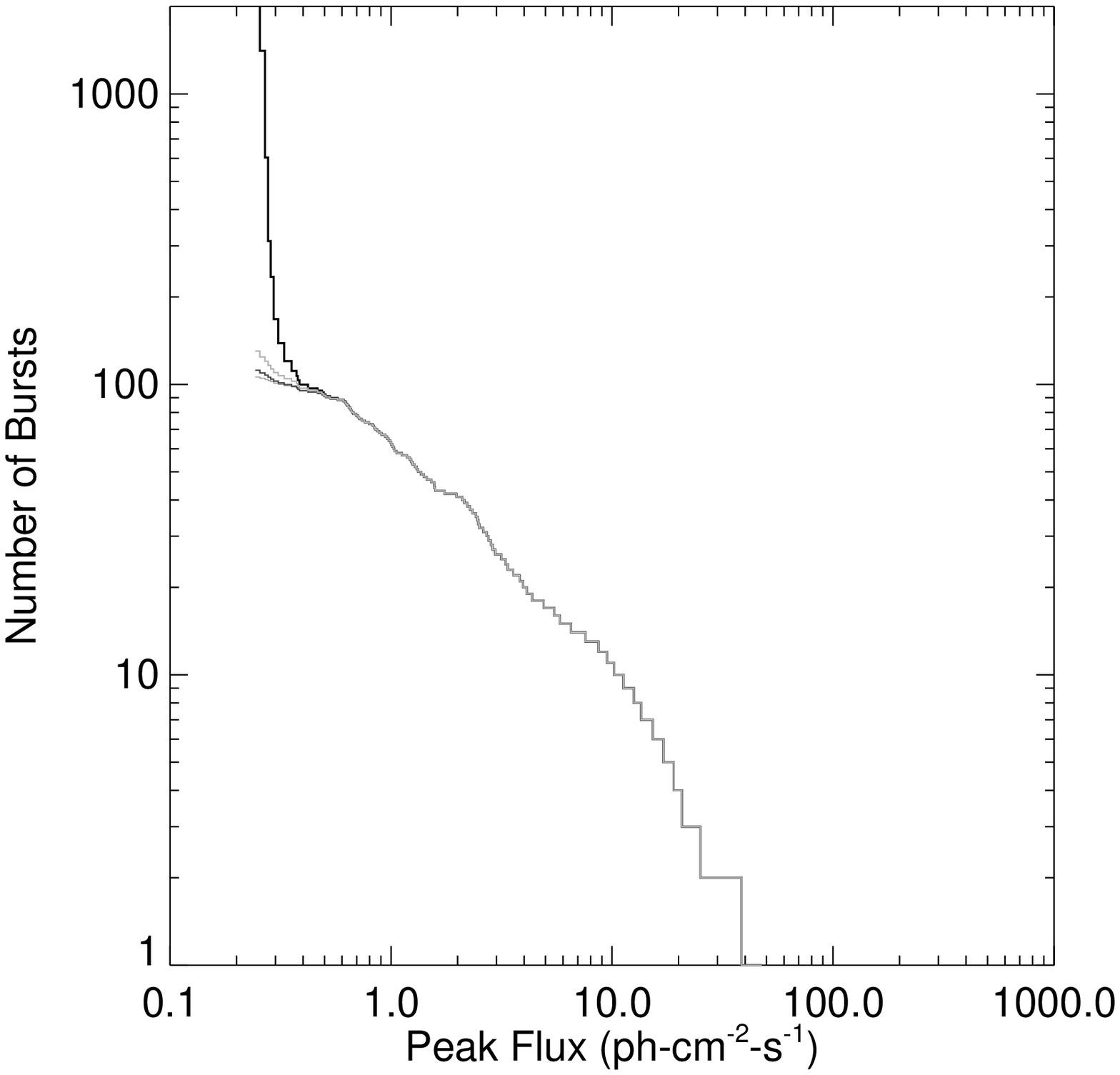,width=6cm,
clip=}
\hfill
\psfig{file=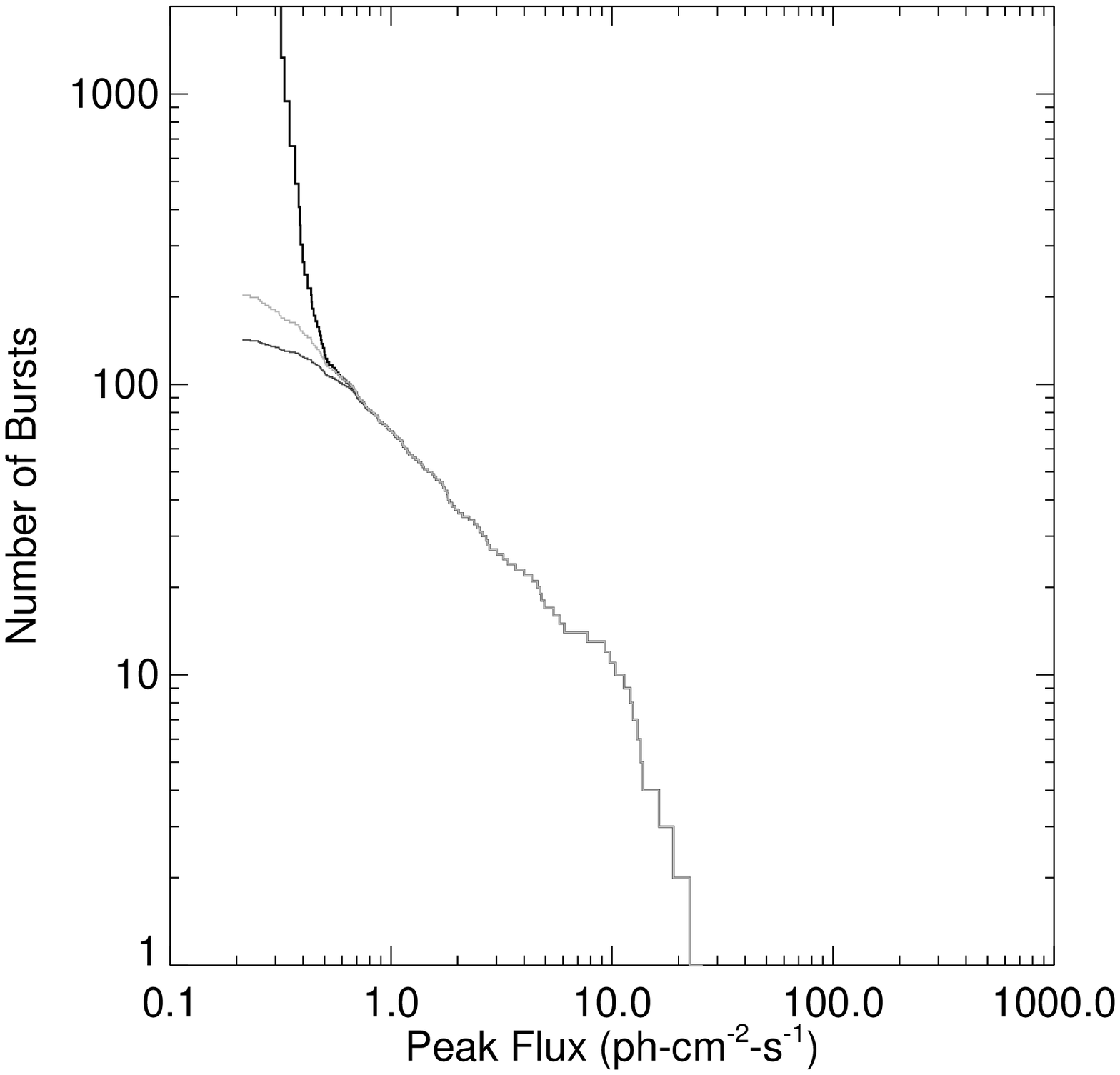,width=6cm,
clip=}}\par\hfill
\caption{\small Integral $\log N - \log P$ distributions for two non-standard
trigger energy ranges: a) 20--100 keV, b) $>$ 100 keV. The peak
flux  $P$ is measured in the standard energy range (50--300 keV).
Near threshold three histograms are shown that represent the
corrections for instrumental trigger efficiency for a range
of incident spectra (see text for details).}
\end{figure*}

The dependence of the instrument sensitivity on the
burst spectrum is further illustrated by the intensity
distribution of bursts detected when the trigger energy
range is different from the nominal. Figure~7a shows the
intensity distribution for trigger energy 25--100 keV; near
threshold, the lowest curve shows the uncorrected data, and
the highest curve results from the assumption that all
bursts have a spectrum with $E_{\rm p}=267$ keV. The second-lowest curve
results from assuming that all burst spectra have
$E_{\rm p} = 50$ keV. The
second-highest curve is the result of combining several
burst spectra with $E_{\rm p} \geq 50$ keV, and it produces an intensity
distribution with an abrupt change in slope near threshold.
Further modeling is necessary in this case. Figure~7b shows
the intensity distribution for the trigger energy range $>$
100 keV. Again, the lowest curve shows the uncorrected data.
The highest curve assumes an input spectrum with $E_{\rm p}=147$ keV.
Assumption of an input spectrum with $E_{\rm p}=1391$ keV results in a
negligible correction. In contrast to Figure~7a, the
intermediate curve, which is an average of several spectra,
appears to produce a reasonably smooth correction.

In summary, the burst intensity distributions collected
in the three different trigger energy ranges each measure a
different component of the burst spectral distribution near
threshold, and deconvolution of the threshold effects is
necessarily model dependent.

\subsection{Durations}

     As with the previous catalogs, we use $T_{50}$ and $T_{90}$ as measures
of burst duration.  $T_{50}$ is the time interval in which the
integrated counts from the burst increases from 25\% to 75\%
of the total counts; $T_{90}$ is similarly defined. Figure 8 shows
the $T_{50}$ and $T_{90}$  distributions for all bursts of the 4Br catalog,
and separately for three different trigger energy ranges.
The well-known bimodality (Kouveliotou et al.\ 1993) is
clearly evident in the overall data and the 50--300 keV data.
The distributions for the lower and higher trigger energy
ranges are consistent, within their limited statistical
accuracy, with the 50--300 keV distribution.

\section{Summary}

     The 4Br catalog includes 1637 cosmic gamma-ray bursts
detected by BATSE during more than five years of operation.
After processing with the most up-to-date location
algorithm, the sky distribution of the bursts remains highly
isotropic. Samples of bursts obtained with trigger energy
ranges different from the BATSE nominal (50--300 keV) are,
within the limited statistics, also isotropic and
inhomogeneous, and have similar duration distributions.

\acknowledgements

The authors gratefully acknowledge the efforts of the
BATSE Operations Team in processing the 4Br catalog data,
and the assistance of B. Schuft, P. Welti, X. Chen, M.
Ivanushkina, and L. Raschke in producing the trigger
efficiencies. Support for UAH \& USRA personnel was provided
through cooperative agreement NCC 8-65. JH was supported
through grant NAG 5-3591. JPL was supported through
cooperative agreement NCC 8-82. Additional support for
summer students was provided through the NSF Research
Experience for Undergraduates program at UAH.

\noindent\parbox{8.75cm}{
\hfil\psfig{file=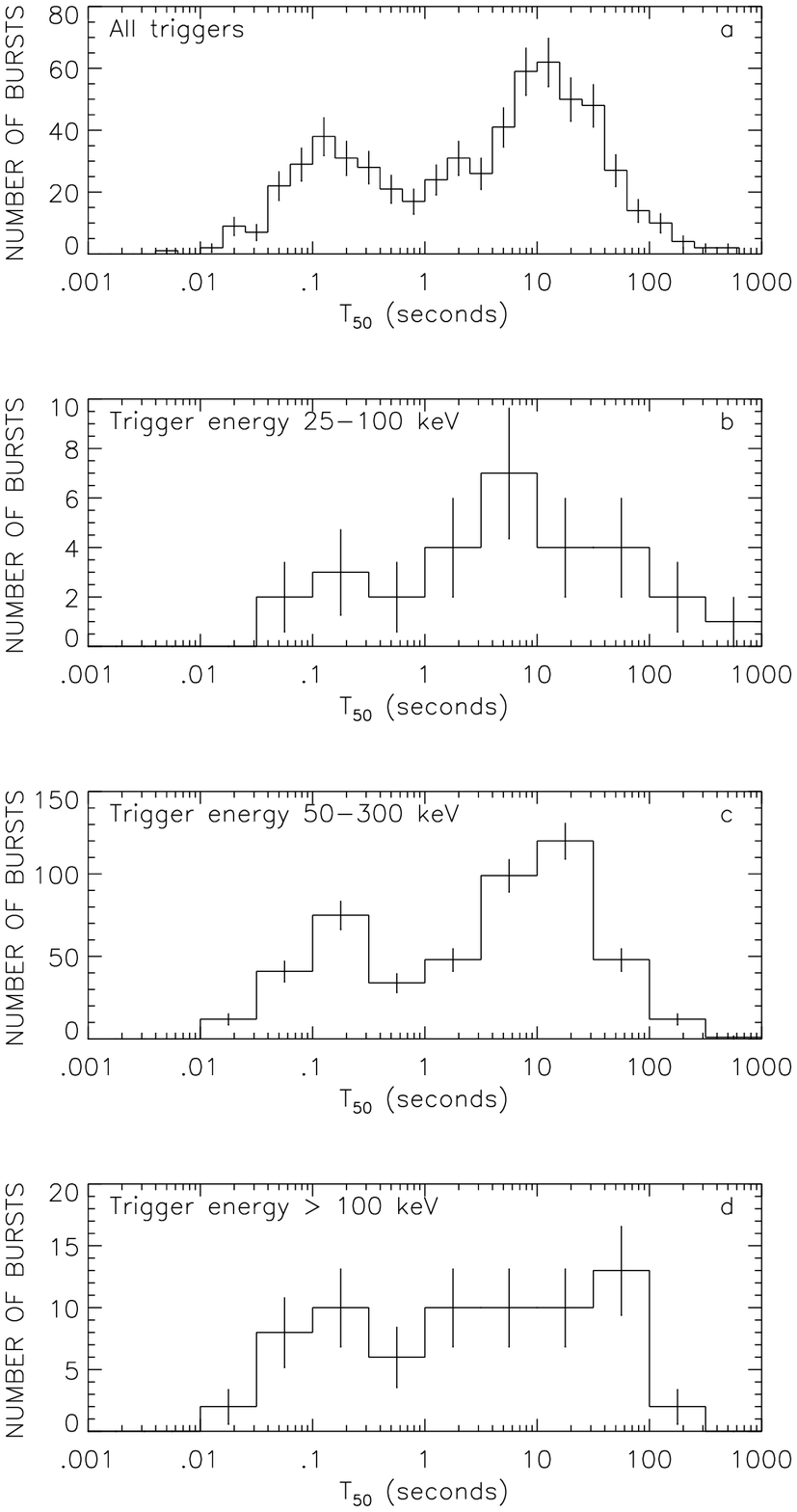,height=12cm,clip=}\hfill
\figcaption{Duration distributions ($T_{50}$) for bursts in the 4Br
catalog that do not have significant data gaps and that are
above the 64 ms trigger threshold. The latter selection
reduces the effect of an instrumental cutoff of short
bursts. a) all bursts,  b) trigger energy range 25--100 keV,
c) trigger energy range 50--300 keV, d) trigger energy range
$>$ 100 keV.}
\vspace{15pt}
}


\end{document}